\documentstyle[prl,aps,preprint]{revtex}

\begin{document}
\tighten
\title{GLUON SPIN IN THE NUCLEON \thanks
{This work is supported in part by funds provided by the U.S.
Department of Energy (D.O.E.) under cooperative
agreement \#DF-FC02-94ER40818 and \#DE-FG02-92ER40702
and in
part by funds provided by the National Science Foundation under grant
\# PHY 92-18167}}

\author{R. L. Jaffe}

\address{Center for Theoretical Physics \\
Laboratory for Nuclear Science \\
and Department of Physics \\
Massachusetts Institute of Technology \\
Cambridge, Massachusetts 02139 \\
and\\
Lyman Physics Laboratory\\
Harvard University\\
Cambridge, Massachusetts 02138 \\
{~}}

\date{MIT-CTP-2466\footnote{Revised Version}  HUTP-95/A034 ~~~ To be
published in {\it Physics Letters B} ~~~ October 1995}
\maketitle

\begin{abstract}

We study the operator description of the gluon spin contribution
($\Gamma$) to the nucleon's spin as it is measured in deep inelastic
processes.  $\Gamma$ can be related to the forward matrix element of a
local gluon operator in $A^+=0$ gauge.  In quark models the nucleon
contains ambient color electric and magnetic fields.  The latter are
thought to be responsible for spin splittings among the light baryons.
We show that these fields give rise to a significant {\it negative\/}
contribution to $\Gamma$ at the quark model renormalization scale,
$\mu_0^2$.  The non-Abelian character of QCD is responsible for the sign
of $\Gamma$.  In a generic non-relativistic quark model
$\Gamma_{NQM}=-{8\over 9}{\alpha_{NQM}\over m_q}\langle{1\over
r}\rangle$, in the bag model $\Gamma_{bag}=-.1\alpha_{bag}$.  These
correspond to $\Gamma_{NQM}\approx -0.7$ and $\Gamma_{bag}\approx -0.4$
at $\alpha_{QCD}\approx 1.0$.

\end{abstract}

\break
\section{Introduction}
Since quark spin accounts for only a small fraction, $\Sigma\approx 0.3$, of
the nucleon's spin$^{\cite{EMC87,Rev95}}$ one wonders where the rest of the
spin resides.  Sehgal$^{\cite{Seh74}}$ pointed
out long ago that experimental data on
hyperon $\beta$-decays and the assumption
that any strange quarks in the nucleon are
unpolarized$^{\cite{EJ74}}$ requires $\Sigma\approx0.6$ and he suggested that
quark {\it orbital\/} angular momentum ($L_Q$)
-- expected in relativistic quark
models -- was a likely candidate.

More recently, the possibility that gluons may carry a significant
fraction of the nucleon's spin has been raised and
debated.$^{\cite{AR,CCM,JM90,BQ90}}$  Much
debate has centered on the possibility that a
gluonic contribution may ``contaminate''
the axial charge sum rules used to extract
$\Sigma$ from polarized lepton scattering data, leading some to speculate that
$\Gamma$ is large and positive.  Whatever their impact on
axial charges, there is no doubt
that gluon spin ($\Gamma$)
and orbital angular momentum ($L_G$) can contribute to the nucleon spin.  In
QCD the nucleon's total angular momentum can be written as
\begin{equation}
{1\over 2} = L_Q +L_G +{1\over2} \Sigma + \Gamma.
\label{eq:0}
\end{equation}
The definition of $\Sigma$ and its
relation to the triangle anomaly were explored shortly after the
publication of the SMC data.$^{\cite{AR,CCM,JM90,BQ90}}$  However,
the definitions of the other terms and their
relation to local operators have received less attention.
No one knows how to measure $L_Q$ and
$L_G$.  $\Gamma$ is defined as the integral of the polarized gluon
distribution,
\begin{equation}
\Gamma(Q^2)=\int_0^1 dx \,\Delta g(x,Q^2) = \int_0^1 dx
(g_\uparrow(x,Q^2)-g_\downarrow(x,Q^2))
\label{eq:1}
\end{equation}
in the sense of a parton model sum rule.

The purpose of this Letter is to study $\Gamma$.
First we review the definition of $\Delta g$ in terms of gauge invariant
operators.$^{\cite {Man91}}$  Next we integrate $\Delta g$, and
specialize to $A^+=0$ gauge, where $\Gamma$ can be related to the matrix
element of a {\it local\/} gluon operator.
We discuss the gauge dependence of $\Gamma$ and its relation
to the spin generators obtained via Noether's theorem.
We conclude that
$\Gamma$ is given by the matrix element of products of gluon vector potentials
and field strengths {\it in the nucleon rest frame and in $A^+=0$ gauge\/},
\begin{equation}
\Gamma(Q^2)={1\over 2M}\langle \hat e_3\vert\,2\,{\rm Tr}\{(\vec
E\times\vec  A)^3 + \vec A_\perp\cdot\vec
B_\perp\}\Big\vert_{Q^2}\vert\hat e_3\rangle
\label{eq:2}
\end{equation}
where $\perp$ refers to the directions transverse to the $\hat e_3$ axis
defined by the target spin, and
$Q^2$ is the renormalization scale of the operators.\footnote{An
earlier version of this paper contained an incorrect version of this equation
which was incorrectly claimed to be gauge invariant.
The author wishes to thank Ian Balitsky and Xiangdong Ji for
raising questions of gauge invariance and for helpful discussions on other
issues.}

Eq.~(\ref{eq:2}) is amenable to evaluation in constituent quark models.
Just as quark models make predictions for $\Sigma$ which are tied to the
phenomenology of $\beta$-decay axial currents, so too they make predictions
for $\Gamma$, tied to the phenomenology of baryon mass differences.
Quark models claim that baryon spin
splittings ({\it e.g.\/} $M_\Delta - M_N$) originate in lowest order exchange
of transverse (magnetic) gluons.  The predictions agree well with
experiment, and the presentations
have found their way into textbooks.$^{\cite{DGH,Bad}}$
Known as ``color magnetism'', this effect
implicitly requires ambient color magnetic fields within hadrons.
The model calculations are performed to lowest
non-trivial order in $\alpha_{QCD}$, where the gluons behave like eight
Abelian vector fields coupled to color.
The spin dependent piece of the Born graph for gluon exchange
between quarks $i$ and $j$ can be rewritten as
\begin{equation}
\Delta M = -\sum_{i>j}\sum_{a=1}^8 \int d^3x\vec
B^a_i(\vec x)\cdot\vec B^a_j(\vec x)
\label{eq:3}
\end{equation}
just as in electrodynamics.  Thus the baryons contain color magnetic (and
electric) fields which may carry spin angular momentum.
The sign of $\Gamma$ is
correlated with the sign of baryon spin splittings and originates in
the non-abelian character of QCD forces.
If quark spin forces were abelian, the $\Delta$
would be {\it lighter\/} than the nucleon and $\Gamma$ would be positive.
The magnitude of $\Gamma$ depends on
details of quark model wavefunctions and
on the renormaliation scale assigned to the quark
model calculations.  However the model's
prediction of the $\Delta - N$ mass difference
constrains the size of $\Gamma$ to some extent.

A generic non-relativistic quark model (NQM) yields $\Gamma_{NQM}=-{8\over
9}{\alpha_{NQM}\over m_q}\langle{1\over r}\rangle$; the standard bag model
yields $\Gamma_{bag}=-0.1\alpha_{bag}$.
Note that $2\Gamma$ is the gluon spin fraction of the
nucleon spin.  We hesitate to assign a specific value to $\alpha$
for fear the resulting number will be taken too seriously.  On the other
hand the model parameters are constrained to some extent by the magnitude of
baryon spin splittings making a numerical estimate possible.
In the non-relativistic quark model the
$N-\Delta$ splitting is $\Delta M = {8\pi\alpha_{NQM}\over
{3m_q^2}}\langle\delta^3(\vec r)\rangle$.  If, for example, we choose Gaussian
wave-functions scaled to reproduce the proton's charge radius, and a quark
mass $m_q\approx M_N/3$ to obtain approximately correct magnetic moments, then
$\Delta M \approx  0.35\alpha_{NQM}M_N$, whence
$\alpha_{NQM}\approx 0.9$.  With this value of
$\alpha_{NQM}$ we find $\Gamma_{NQM}\approx -0.8$.
In the bag model,
$\alpha_{bag}\approx 2$
in order to fit baryon mass differences, whence $\Gamma_{bag}\approx -0.2$.

Note that these estimates
apply at the {\it quark model renormalization
scale\/}.  To obtain a prediction for $Q^2\sim
2-10\,GeV^2$ relevant to experiment,
it is necessary consider the evolution of $\Gamma$ with
renormalization scale.  It has long been known that $\Gamma$ evolves
homogeneously, and that the quantity $\alpha(Q^2)\Gamma(Q^2)$
is a renormalization group
invariant to leading order.$^{\cite{RG}}$  Therefore the sign of $\Gamma$
should be a reliable prediction of the models.  If we take the numerical
estimates seriously, evolution will tend to bring the two predictions closer
together because the bag estimate will
be multiplied by a larger factor
of $\alpha(\mu_0^2)/\alpha(Q^2)$.
For example, at $\alpha_{QCD}\approx 1$, $\Gamma_{bag}\approx- 0.4$, and
$\Gamma_{NQM}\approx -0.7$.

In the next section we introduce and analyze the operator
measure of gluon spin.  We discuss gauge invariance.
We show that the operator we evaluate
in the rest frame is the same one measured
by integrating the gluon helicity asymmetry
$\Delta g(x,Q^2)$.  In Section 3, we estimate the
magnitude of $\Gamma$ in simple models.
In Section 4
we discuss the reliability of our calculation and
and mention further applications.

\section{The Operator Description of Gluon Spin}

Experimenters will not measure $\langle(\vec
E\times\vec A)^3 +\vec A_\perp\cdot\vec B_\perp\rangle$ directly.  Instead they
will measure the polarized gluon distribution
function, $\Delta g(x,Q^2)$, in deep
inelastic lepton scattering.  $\Delta g$ measures
the probability to find a gluon with its
helicity parallel to the nucleon's helicity minus the
probability to find it antiparallel.  The integral of $\Delta g$
measures
$\Gamma$ and therefore
will tell us which local operator, if any, is to be associated
with the gluon spin.  We begin with the operator representation
of the polarized
gluon distribution function,$^{\cite{Man91}}$
\begin{eqnarray}
\Delta g(x,Q^2)&=&{i\over 4x\pi P^+}\int
d\xi^- e^{-ix\xi^-P^+}\langle P,\hat e_3\vert
\,{\rm Tr}\,\{F^{+\alpha}(\xi^-,\vec 0){\cal I}(\xi^-,0)\tilde
F_\alpha^{\phantom{\alpha}+}(0)\}\Big\vert_{Q^2}\vert P,\hat
e_3\rangle \nonumber \\
&+&(x\rightarrow -x),
\label{eq:5}
\end{eqnarray}
where $F$ (and $A$) are matrices
($F\equiv\sum_{a=1}^8F^a\lambda^a$, {\it
etc.\/}).  $\{F^a\}$ are in the adjoint, and
$\{\lambda^a\}$ are in the triplet representation (with
${\rm Tr}\{\lambda^a\lambda^b\}={1\over 2}\delta^{ab}$).
$\xi^\pm,\vec\xi^\perp$ are light-cone coordinates
and $(\xi^-,\vec 0)$ denotes the point
$\xi^-, \xi^+=\vec\xi^\perp = 0$.  The label
$Q^2$ is a reminder that the tower of local matrix
elements in the Taylor expansion of $F\tilde
F$ are understood to be renormalized at a
factorization scale, $Q^2$, and finally ${\cal I}$ is the Wilson-line integral,
\begin{equation}
{\cal I}(\xi^-,0)={\cal P}{\rm exp}\left(ig\int_0^{\xi^-}dy^-A^+(y^-,\vec
0)\right).
\label{eq:6}
\end{equation}
The standard parton interpretation follows from
eq.~(\ref{eq:5})  if we choose $A^+=0$
gauge and introduce the momentum decomposition of the fields $F$ and $\tilde
F$ quantized at $\xi^+=0$.

In order to integrate
eq.~(\ref{eq:5}) over $x$ we must study the apparent singularity at
$x=0$.  Physically, $\Delta g(x,Q^2)$ is not expected to
diverge as fast as $1/x$, so the $\xi^-$--integral must vanish as
$x\rightarrow 0$.  This means that the singularity at $x=0$ is integrable.
If $\Delta g$ were found to
diverge like $1/x$ or faster, our analysis would have to be reconsidered.
With this in mind, we can interchange the $x$ and $\xi^-$ integrations and,
because $\Delta g$ is symmetric in $x$, we can use the principal value
prescription at $x=0$,
$\int\hspace{-3.5mm}-{dx\over x} e^{-i\alpha
x}=-i\pi\varepsilon(\alpha)$.  We obtain
\begin{equation}
\Gamma(Q^2)={1\over 2P^+}\int d\xi^-\varepsilon(\xi^-)\langle P,\hat
e_3\vert\,{\rm Tr}\,\{F^{+\alpha}(\xi^-){\cal I}(\xi^-,0)
\tilde F_{\alpha}^{\phantom{\alpha}+}(0)\}\Big\vert_{Q^2}\vert
P,\hat e_3\rangle.
\label{eq:6a}
\end{equation}
This expression cannot be simplified further unless we choose $A^+=0$ gauge.

In $A^+=0$ gauge ${\cal I}=1$ and $F^{+\alpha}={\partial\over\partial
\xi^-}A^\alpha$, so we may perform the $\xi^-$ integration.  The terms at
$\xi^-=\pm\infty$ vanish because
the integral in eq.~(\ref{eq:5}) converges when $x=0$.
Only the local ($\xi^-=0$) contribution
survives.
We choose the rest frame for
$P$, and after some algebra we are left with the expression we seek,
\begin{eqnarray}
\Gamma(Q^2)&=& {1\over \sqrt{2}M}\langle P, \hat e_3\vert 2\,{\rm
Tr}\,\{A^1F^{+2}-A^2F^{+1}\}\Big\vert_{Q^2}\vert P,\hat e_3\rangle
\nonumber\\
&=& {1\over{ 2M}}\langle
P,\hat e_3\vert 2\,{\rm Tr}\, \{ (\vec E\times\vec
A)^3 +\vec A_\perp\cdot\vec B_\perp\}\Big\vert_{Q^2}\vert P,\hat e_3\rangle,
\label{eq:7}
\end{eqnarray}
where $E^i=F^{i0}$, and $B^i=-{1\over 2}\varepsilon^{ijk}F^{jk}$.  The
choice of $A^+=0$ gauge was essential to this derivation.  Without it,
$\Gamma$ does not appear to be associated with a {\it local\/} operator.

$A^+=0$ does not completely fix the gauge in QCD.  Residual
non-abelian gauge transformations, $\delta A^\mu = \partial^\mu\delta\alpha
+[\delta\alpha,A^\mu]$, are allowed provided they
obey $\partial^+\delta\alpha=0$ which preserves $A^+=0$.
$\Gamma$ must be invariant under this residual gauge symmetry.  Using the
Bianchi identity --- $[D^\mu,F^{\nu\lambda}] + [D^\nu,F^{\lambda\mu}]
+[D^\lambda,F^{\mu\nu}]=0$ --- it is straightforward to show that
\begin{equation}
\delta\Gamma\propto\langle P,\hat e_3\vert {\rm Tr}\,\{\alpha\partial^+
F^{12}\}\vert P,\hat e_3\rangle
\label{eq:7a}
\end{equation}
which vanishes because $\alpha$ is independent of $\xi^-$ and the
$+$ derivative sits inside a forward matrix element.  Thus $\Gamma$ is
invariant under the residual gauge symmetries of $A^+=0$ gauge.

Eq.~(\ref{eq:7}) is not a familiar representation for the spin angular momentum
stored in a gauge vector field.  Happily it can be related directly to the
gluon spin term in the angular momentum tensor density in QCD.
Angular momentum
in QCD, as in any field theory, is described by a rank-$3$ Lorentz tensor,
$M^{\mu\nu\lambda}$.  In a physical gauge, where there are no ghosts,
\begin{eqnarray}
M_{QCD}^{\mu\nu\lambda}&=&
{i\over 2}\bar\psi\gamma^\mu
\left(x^\lambda\partial^\nu-x^\nu\partial^\lambda\right)\psi
+{1\over 2}\epsilon^{\mu\nu\lambda\sigma}
\bar\psi\gamma_\sigma\gamma_5\psi \nonumber \\
&-&2\,{\rm Tr}\,\{F^{\mu\alpha}
\left(x^\nu\partial^\lambda-x^\lambda\partial^\nu\right)
A_\alpha\}
+ 2\,{\rm Tr}\,\{F^{\mu\lambda}A^\nu+F^{\nu\mu}A^\lambda\} \nonumber \\
&-&{1\over 2}\,{\rm Tr}\, F^2
\left(x^\nu g^{\mu\lambda}-x^\lambda g^{\mu\nu}\right),
\label{eq:4}
\end{eqnarray}
The second term in eq.~(\ref{eq:4}) measures the quark spin -- at least up
to subtleties arising from the triangle anomaly.$^{\cite{AR,CCM,JM90,BQ90}}$
The first and third terms look like
the quark and gluon orbital angular momentum
respectively, because they have the standard ``convective'' form of orbital
angular momentum in a field theory, $\Pi^\dagger(\vec
x\times\vec\nabla)\Phi$, where $\Pi$ and $\Phi$ are canonical coordinate and
momentum respectively.  The last term contributes only to boosts.  The fourth
term is a candidate for the gluon spin.
\footnote{Eq.~(\ref{eq:4}) was derived in Ref.~\cite{JM90}, where the
generator of gluon spin rotations, $M^{0ij}_\Gamma$ was incorrectly identified
as $\vec A\times\vec E$.  Since this sign is crucial to our results we have
checked a) that canonical tranformations give $\vec E\times\vec A$ and b) that
the two gluon contributions combine to give a total angular momentum of $\vec
J = \vec x\times\vec E\times\vec B$ as expected.}
Let us define,
\begin{equation}
M_\Gamma^{\mu\nu\lambda}\equiv 2\, {\rm Tr}
\{F^{\mu\nu}A^\lambda+F^{\lambda\mu}A^\nu\}.
\label{eq:3a}
\end{equation}
Then comparison with eq.~(\ref{eq:7}) shows that
\begin{equation}
\Gamma(Q^2)={1\over 2S^+}\langle P,\hat e_3\vert
M^{+12}_{\Gamma}\Big\vert_{Q^2}\vert P,\hat e_3\rangle,
\label{eq:3b}
\end{equation}
in $A^+=0$ gauge.  This identification makes physical sense since the parton
model distribution should measure helicity along the $\hat e_3$-axis
(hence $\nu=1$, $\lambda=2$) in an infinite momentum frame, which corresponds
to $\mu=+$ in the laboratory.  The restriction to $A^+=0$ gauge is natural in
the parton model.  Needless to say, this restriction does not render $\Gamma$
gauge dependent:  There is a corresponding operator definition of $\Gamma$ in
any gauge.  However it is not simple or even local.
Note that this definition of gluon spin {\it does not\/} correspond to the
gluon piece of the generator of rotations in the laboratory, which would be
$M_\Gamma^{012}$ and would not naturally appear in light-cone gauge.

This discussion suggests that a natural definition of the quark and gluon
{\it orbital angular momentum\/} might select the $+12$
component of the appropriate piece of $M^{\mu\nu\lambda}$
in eq.~(\ref{eq:4}).$^{\cite{Ji95}}$
While this
is attractive, a physically
interesting $L_Q$ or $L_G$ must be determined by what can be measured
experimentally.  As long as no measure of orbital angular momentum
is experimentally accessible, the definition will remain open to
question.

\section {The Gluon Spin in Quark Models}

Quark models of the light hadrons fall into
two general classes:  non-relativistic quark
models, where quarks are described by the
Schroedinger equation (perhaps including
relativistic corrections) and confined by some
two body color dependent forces; and bag
models, where relativistic quarks, governed by
the Dirac equation, move in some confining background field imagined to be
self-consistently generated by their deformation
of the non-perturbative QCD vacuum.
Both extremes give good explanations
of the mass spectrum of the lightest hadrons
(pseudoscalar and vector mesons, and octet
and decuplet baryons).  A major role is played by color mediated, spin
dependent forces.$^{\cite{DGG75,DJJK75}}$  In this
Section we shall see that the gluons responsible
for these spin splittings are {\it anti-aligned\/} with
the nucleon spin ($\Gamma<0$);
that this is a particular consequence of the
non-abelian nature of QCD interactions, and
that the effect has roughly the same
magnitude in both types of model.

The quantity we
want to evaluate in models is\footnote{
Quark model states are more conveniently normalized to
unity than covariantly, so we change normalization accordingly,
$\langle P\vert{\cal O}(0)\vert P\rangle= 2M\int d^3x \langle T\vert{\cal
O}(\vec x)\vert T\rangle$,
where $\vert T\rangle$ is a quark model state normalized to unity.}
\begin{equation}
\Gamma(\mu_0^2)=\langle T,\hat e_3\vert\int d^3x \,2{\rm Tr}\,\{
\left(\vec E(\vec x)\times\vec A(\vec
x)\right)^3+\vec A_\perp(\vec x)\cdot\vec B_\perp(\vec x)\}
\vert T,\hat e_3\rangle
\label{eq:8a}
\end{equation}
One way to evaluate this expression would be to compute the relevant Feynman
diagram --- obtained by inserting this gluonic operator into the lowest order
Born diagram for gluon exchange between bound quarks (in $A^+=0$ gauge).
The graphical method would require us to
construct and use {\it confined\/} gluon
Green's functions in $A^+=0$ gauge, which is
unnecessarily complicated.  It is
easier to compute the ambient color fields by directly integrating the QCD
equations of motion (which reduce to eight
copies of Maxwell's equations in the
Abelian approximation).  In this way we obtain expressions for $\vec A$, $\vec
B$, and $\vec E$ which depend on the quark color, spin and spatial coordinates
and can be evaluated with the help of model wavefunctions.

In a generic quark model the color-electric fields will be given by the
gradient of a time-independent function of the quark degrees of freedom,
$\vec E^a(\vec x) = -\vec\nabla \Phi^a(\vec x)$, with
\begin{equation}
\Phi^a(\vec x) = {g\over 4\pi}\sum_i \lambda_i^a
{\cal G}(\vec x,\vec x_i)
\label{eq:9}
\end{equation}
$\Phi$ is an operator in the space of the quark color ($\lambda^a_i$) and
position ($\vec x_i$) states.   It depends on
some model Green's function ${\cal G}$. For example, in an unconfined,
non-relativistic model ${\cal G} = 1/\vert \vec x -\vec x_i\vert$.

The magnetic field, $\vec B^a(\vec x)$ can likewise be written as the $curl$ of
a time independent function of the quark variables, $\vec B^a(\vec
x)=\vec\nabla\times\vec U^a(\vec x)$, with
\begin{equation}
\vec U^a(\vec x)={g\over 4\pi}\sum_i\lambda^a_i\vec\sigma_i\times\vec{\cal
G}(\vec x,\vec  x_i).
\label{eq:10}
\end{equation}
For example, $\vec{\cal G}=(\vec x - \vec x_i)/2 m_q \vert
\vec x -\vec x_i\vert^3$ in a non-relativistic model (where $\vec m^a_i\equiv
g\vec\sigma_i\lambda^a_i/2m_q$ is the quark's color magnetic moment operator).
The nucleon polarization (and the gauge choice, $A^+=0$) selects the $\hat
e_3$--axis.  Matrix elements of the operator
$\vec\sigma_i$ will therefore vanish
except in the $\hat e_3$ direction.  So we conclude that $\langle\vec
U^a_i\cdot\hat e_3\rangle =\langle U^{a3}_i\rangle = 0$.

So far this result is quite general.  It holds in any quark model where the
gluons are treated to lowest order (abelian approximation) and
$\vec U$ has no component along the nucleon spin.  In
``symmetric'' quark models the quark distributions in the nucleon ground state
are not correlated with the overall spin.  When
integrated over the wavefunction of the
$j^{th}$ quark, then, the resulting color electric field is radial and
independent of $j$,
\begin{equation}
-\int d^3x_j
\vec\nabla{\cal G}(\vec x,\vec x_j)\vert\psi(\vec x_j)\vert^2 = {\vec x\over
r^3}Q(r),
\label{eq:16b}
\end{equation}
where $r=\vert\vec x\vert$ and $Q(r)=4\pi\int_0^rdr' r'^2
\vert\psi(r')\vert^2$ is the color charge inside the sphere with radius
$r$.  Likewise integration over $\vec x_i$ simplifies $\vec {\cal G}$,
\begin{equation}
\int d^3x_i \vec{\cal G}(\vec x,\vec x_i)\vert\psi(\vec x_i)\vert^2 = \vec x
h(r),
\label{eq:16c}
\end{equation}
where h(r) describes the vector potential generated by the model dependent
magnetization density.

The operators $\Phi$ and $\vec U$ are not yet the appropriate
scalar and vector potentials for the
gluon field because they do not satisfy the $A^+=0$ gauge condition.
However suitable potentials are easily constructed.
Define
\begin{eqnarray}
A^{0a}(\vec x)&=&\Phi^a(\vec x)\nonumber\\
\vec A^a(\vec x)&=&\vec U^a(\vec x)-\vec\nabla\int_0^z d\zeta\Phi^a(x,y,\zeta)
\end{eqnarray}
These potentials generate $\vec E^a$ and $\vec B^a$ in the usual way and
satisfy the gauge constraint, $A^{0a}+A^{3a}=0$ (remembering $U^{3a}=0$).
The choice of time independent potentials as well as the lower limit on the
$\zeta$ integration correspond to residual gauge freedom available in $A^+=0$
gauge.

We now substitute the operator definitions of $A^{0a}$ and $\vec A^a$
into  eq.~(\ref{eq:8a}) and obtain
\begin{eqnarray}
\Gamma(\mu_0^2)&=&\sum_{i\ne j}\sum_{a=1}^8\int d^3x\langle T, \hat e_3\vert
\{ \left[\vec E^a_i(\vec
x)\times\vec U^a_j(\vec x)\right]^3 + \left[\vec E^a_i(\vec x)\times
\vec\nabla f^a_j(\vec x)\right]^3\nonumber\\
&+&\vec U^a_{\perp i}(\vec x)\cdot \vec B^a_{\perp j}(\vec x)
+\vec\nabla_\perp f^a_i(\vec x)\cdot\vec B^a_{\perp j}(\vec x)\}\vert T, \hat
e_3\rangle
\label{eq:14}
\end{eqnarray}
where we have separated out the contributions from individual quarks
$i$ and $j$ to each of the field operators.  Note we have dropped the
$i=j$ terms.  These correspond to ``self-angular momentum'' effects and are
associated with questions of renormalization, which are discussed further in
Section 4.

The second and third terms in eq.~(\ref{eq:14}) vanish.  The
second vanishes after integration by parts because $\vec\nabla\times\vec E^a_j
= 0$.  There is no associated surface term.  The third term vanishes for
spatially symmetric quark wavefunctions.

Consider now the fourth term and write out the space components explicitly
(suppressing the color ($a$) and quark ($i,j$) labels, and bras and kets,
\begin{eqnarray}
\Gamma_4&=&\int d^3x \left({\partial f\over\partial y}
{\partial U_1\over\partial z} - {\partial f\over\partial x}
{\partial U_2\over\partial z}\right)\nonumber\\
&=&\int d^3x \left(E_1U_2-E_2U_1\right)\nonumber\\
&+&\oint_{S_R}d^2s\hat e_3\cdot\hat r\left(
U_1(\vec x)\int_0^zd\zeta E_2(x,y,\zeta)
-U_2(\vec x)\int_0^zd\zeta E_1(x,y,\zeta)\right)
\label{eq:15}
\end{eqnarray}
where the surface integral is over a sphere at large distance (for unconfined
gluons) or the bag surface (for bag-like models).  The first term in
eq.~(\ref{eq:15}) is identical to the first term in eq.~(\ref{eq:14}).

We combine these results and substitute
the parameterizations of $\vec E$ and
$\vec U$ from eq's.~(\ref{eq:16b}) and (\ref{eq:16c})
to obtain,
\begin{equation}
\Gamma={8\over 9}\alpha \int_0^RdrrQ(r)(h(R)-2h(r)),
\label{eq:18}
\end{equation}
where the $r$-integration goes to infinity in generic non-relativistic quark
models, but ends at $r=R$, the bag surface in the bag model.  The term
proportional to $-2h(r)$ is the volume integral of $\vec E\times\vec U$, and
the $h(R)$-term is the surface contribution left over from integration by
parts.  In reaching eq.~(\ref{eq:18}) we have used
$\sum_{a=1}^8\lambda_i^a
\lambda_j^a=-2/3$ for $i\ne j$.  [Because the nucleon
is a color singlet, $[\sum_{a=1}^8\sum_{j=1}^3\lambda_j^a]^2=0$.  The result
follows since
$\sum_{a=1}^8 (\lambda_j^a)^2=4/3$.]  Also, $\alpha=g^2/4\pi$ and
$\sum_{i\ne j} \langle\sigma^3_i\rangle=2$ in a state polarized along the $\hat
e_3$-direction.

Now let us specialize to quark potential models where
quarks are confined but color is not.  Gluon
field strengths fall off at large distances like abelian multipoles.
The non-relativistic vector potential, $\vec U = \vec m \times \vec r/r^3$,
corresponds to $h(r)_{NQM}\propto 1/2m_qr^3$.  At large $R$, $Q(R)\rightarrow
1$,  so the surface term in eq.~(\ref{eq:18}) vanishes.  Substituting for
$Q(r)$ in terms of the quark wavefunction, $\psi(r)$ and interchanging
integrations, eq.~(\ref{eq:18}) reduces to
\begin{eqnarray}
\Gamma_{NQM}(\mu_0^2) &=&-{8\over 9m_q}\alpha_{NQM}(\mu_0^2)\int_0^\infty dr r
\vert\psi(r)\vert^2\nonumber\\
&=&-{8\over 9m_q}\alpha_{NQM} (\mu_0^2)\langle {1\over r}\rangle,
\label{eq:22}
\end{eqnarray}
where we have restored the quark model renormalization scale, $\mu_0^2$, to
remind us that this value pertains to some low scale at which the model is
formulated.

The parameters $m_q$, $\langle{1\over r}\rangle$
and  $\alpha_{NQM}(\mu_0^2)$ are
all model dependent, but not unconstrained.
$m_q\approx 0.3 GeV$ reproduces nucleon magnetic
moments.  Another constraint comes from the
$\Delta-N$ mass difference which is given by
\begin{equation}
\Delta M = {8\pi\over 3}{\alpha_{NQM}(\mu_0^2)\over
m_q^2}\langle\delta^3(\vec r)\rangle
\label{eq:14a}
\end{equation}
in the non-relativistic quark model.$^{\cite{DGG75}}$
To obtain a numerical estimate, we assume a
gaussian wavefunction adjusted to reproduce
the root-mean-square charge radius of the proton.
Then we find $\alpha_{NQM}(\mu_0^2)\approx
0.9$, and $\Gamma_{NQM}(\mu_0^2)\approx -0.8$.

Static bag model calculations of hadron spin splittings were carried
out by explicit construction of color electric and magnetic
fields.$^{\cite{DJJK75}}$  We may borrow results from that work to evaluate
eq.~(\ref{eq:18}).  The color magnetic field is calculated from
the QCD generalization of Maxwell's equation,
$\vec\nabla\times\vec B^a=g\psi^\dagger\vec\alpha\lambda^a\psi$
augmented by the boundary
condition $\hat r\times\vec B^a=0$ at $r=R$.  A short calculation yields,
\begin{eqnarray}
Q(r)&=&\int_0^rdr' r'^2\left(f^2(r')+g^2(r')\right) \nonumber \\
h(r)&=&\{{1\over 2}{{\mu(R)}\over R^3}+{{\mu(r)}\over r^3}+\int_r^R
dr'{{\mu'(r')}\over {r'^3}}\},\quad{\rm where} \nonumber \\
\mu(r)&=&\int_0^r dr'{8\pi\over 3}r^3f(r)g(r)
\label{eq:19}
\end{eqnarray}
where $f(r)\propto j_0(x_0r/R)$
and $g(r)\propto j_1(x_0r/R)$.  [$x_0$ is the lowest
solution to the eigenvalue condition $\tan x =
x/1-x\quad(x_0=2.0428)$.]
Substituting explicit wavefunctions we find
\begin{equation}
\Gamma_{bag}(\mu_0^2)= -0.1\alpha_{bag}(\mu_0^2)
\end{equation}
Standard bag model calculations of baryon spin splittings require
$\alpha_{QCD}\approx 2$, so we find $\Gamma_{bag}\sim -0.2$ at the
renormalization scale of the model.

\section{Discussion and Conclusions}

Our calculations force us to conclude that the gluons responsible for
the spin splittings among light baryons are {\it anti-aligned\/} with the spin
of their parent nucleon.  This exacerbates rather than helps resolve the
problem with the nucleon spin, suggesting we look elsewhere for a large
positive contribution to the spin.  Our calculation required two assumptions
beyond the traditional formulation of quark models:  first, that
``self-interaction'' contributions to $\Gamma$ could be ignored, and second,
that the estimate of $\Gamma$ at the quark model scale, $\mu_0^2$ has
something to do with its value at experimentally accessible scales.  Here we
will first look at the reliability of our estimate in the context of the
models. Then we will comment further on the two assumptions and finally
mention some possible extensions of the work.

The troubling sign of $\Gamma$ does not appear to depend on the details of the
models.  The directions of the color electric and magnetic fields are fixed by
the charges and currents which give successful descriptions of a variety
of hadronic phenomena including magnetic moments as well as colormagnetic spin
splittings.  The minus sign arises because the gluons are non-abelian --- the
two spectators of any given quark in the nucleon generate color fields
appropriate to an antiquark.  The same effect flips the sign of baryon spin
splittings.  If quark spin forces were abelian, $\Gamma$ would be positive,
but the $\Delta$ would be lighter than the nucleon.

On the other hand,
the magnitude is of $\Gamma$ quite uncertain.  The value of
$\Gamma(Q^2)$ at scales relevant to experiment varies quadratically with the
assumed value of $\alpha(\mu_0^2)$ taken from fits to baryon spin splittings.
One factor of $\alpha$ comes from the operator itself; the other factor arises
in scaling from $\mu_0^2$ to $Q^2$.
There is reason to believe that symmmetric quark models overestimate the
value of
$\alpha$ necessary to account for spin splittings because they ignore the
correlations color magnetic interactions would introduce into wavefunctions.
We expect that the color magnetism would correlate quark pairs into
color $\bar 3$, spin $0$ diquarks (the most attractive channel) with the
effect of increasing the matrix element and decreasing the value of $\alpha$
necessary to reproduce spin splittings.  An estimate of the size of this
effect goes beyond the scope of this Letter.  However we caution the reader
against taking the calculated magnitude of $\Gamma$ very seriously.
The most important conclusion of our work is the prediction that the sign of
$\Gamma$ should be negative.  If experiment finds $\Gamma$ to be positive, and
if neither self-interactions nor evolution change the model predictions, then
the quark model calculations of baryon spin splittings will have to be
re-evaluated.

Regarding the self interaction terms, the proper prescription in principle is
``calculate all graphs and renormalize''.
Consider the case of an isolated electron in QED: the $\vec E$ and $\vec B$
fields which surround it contribute a log-divergent term to the
expectation of $M^{+12}_\Gamma$.  The same occurs
for the stress tensor (one piece of which is the self-energy).
Renormalization removes these infinities.  Since
$M_{\mu\nu\lambda}=x_\mu T_{\nu\lambda}-x_\nu T_{\mu\lambda}$ (up to a
total derivative) renormalizing $T$ also renormalizes $M$.  If we renormalize
on-shell, the spin is found on the renormalized electron line and the
self-fields are to be ignored.  The same argument would seem to carry over to
the quark model
and dictate that we ignore all terms with $i=j$ in eq.~(\ref{eq:8a}).
However we cannot renormalize a confined quark on
mass-shell, so we cannot assume that all self-field effects can be
renormalized away.
The problem also arises in the bag model where the
calculation and {\it renormalization\/}
of the self-field contribution to $M^{+12}_\Gamma$ would
yield a finite result analogous to the Lamb shift correction
to the self-energy of a bound electron.  This program has actually been
carried out for the self-energy of a quark in a bag.$^{\cite{GHJ}}$
Only a complete treatment of the renormalization of a confined quark would
enable one to decide how much of the self-field should be included in the
normalization of a quark at the scale $\mu_0^2$ and how much should be
attributed to its binding inside the nucleon.  We have, in effect,
{\it assumed\/} that {\it all\/} self-field effects are to be included in the
definition of the renormalized quark operators and omitted from our
calculation of $\Gamma$.  This assumption should be examined more closely; it
is possible that finite parts of the $i=j$ terms make a significant
contribution to $\Gamma$ at the scale $\mu_0^2$.

The renormalization point and scheme dependence of $\Gamma$ has been discussed
extensively.  In $A^+=0$ gauge
it is easy to see that the forward matrix element of the
Kogut-Susskind current,$^{\cite{KS74}}$ $\langle K^+\rangle$, coincides with
our definition of $\Gamma$.  So the studies of the QCD evolution of
the forward matrix elements of $\langle K^+\rangle$$^{\cite{AR,JK95}}$ apply
without alteration to the evolution of $\Gamma$.
To leading order $\alpha_{QCD}(Q^2)\Gamma(Q^2)$ is independent of $Q^2$.
Beyond leading order the $Q^2$ dependence of $\Gamma$ is scheme dependent, and
not of much interest until one understands the renormalization scale and
scheme dependence of quark model calculations.  These problems
notwithstanding, it appears that the
quark model result $\Gamma<0$ is preserved by leading order evolution.

The analysis presented here raises several questions and suggests some avenues
for future study.  First, no doubt, is to understand better the origin of the
result $\Gamma_{NQM}<0$ and its generality.  The light-cone perturbation
theory formalism pioneered by Brodsky and Lepage,$^{\cite{BL}}$
and applied to the parton
structure of positronium by Burkardt,$^{\cite{Bur}}$ seems like the natural
framework in which to explore the result further.  Another direction would be
to use these models to explore the $x$--dependence of the nucleon's polarized
gluon distribution (at the scale $\mu_0^2$), since
the shape of $\Delta g$ is of interest.$^{\cite{RLJ75}}$  These
calculations might serve to guide the application of sophisticated
(next-to-leading order) perturbative QCD fits to spin dependent gluon
distributions.$^{\cite{GRV}}$  Finally we note
that there is much less reason to compute the spin-{\it independent\/} gluon
distribution using these models.  The lowest moment of $g(x,Q^2)$ does not
converge and is not associated with a local operator.  Furthermore models that
take seriously the gluon role in confinement, leave us puzzled whether the
long-range confining field ({\it e.g.\/} the bag itself in the bag model)
should be associated with a (spin-independent) gluon distribution or not.

\section{Acknowledgements}
I would like to thank the organizers and participants in the 1995 Erice School
on Nucleon Spin Structure where this work was conceived, for discussions,
and the Physics Department at Harvard University, where it was performed,
for hospitality.


\begin{references}

\bibitem{EMC87}
J.~Ashman {\it et al.\/} Phys. Lett. {\bf B206}, 364 (1988); Nucl. Phys. {\bf
B328}, 1 (1989).

\bibitem{Rev95}  For a review and summary of experimental data and further
references , see C.~Cavata and V.~Hughes,
eds. {\it Proceedings of the Yale Workshop\/}
(Plenum Press, New York, 1995).

\bibitem{Seh74}
L.~M.~Sehgal, Phys. Rev. {\bf D10}, 1663 (1974).

\bibitem{EJ74}
J.~Ellis and R.~L.~Jaffe,  Phys. Rev. {\bf D9}, 1444 (1974), {\bf D10}, 1669
(1974).

\bibitem{AR}A.~V.~Efremov and O.~V.~Teryaev, Czech.~Hadron Symposium (1988),
302;  G.~Alterelli and G.~G.~Ross,
Phys. Lett. {\bf B212} (1988) 391.

\bibitem{CCM}R.~D.~Carlitz, J.~C.~Collins
and A.~H.~Mueller, Phys. Lett. {\bf B214} (1988) 229.

\bibitem {JM90}R.~L.~Jaffe and A.~Manohar, Nucl. Phys. {\bf B337} (1990) 509;

\bibitem {BQ90}G.~Bodwin and J.~Qiu, Phys. Rev. {\bf D41} (1990) 2755.

\bibitem{Man91}J.~C.~Collins and D.~E.~Soper,
Nucl. Phys. {\bf B194} 445 (1982);
A.~Manohar, Phys. Rev. Lett. {\bf 65} (1990)
2511, {\it ibid.\/} {\bf 66} (1991) 289.

\bibitem {DGH} {\it Dynamics of the Standard Model\/} By
J.~F.~Donoghue, E.~Golowich, and B.~R.~Holstein, (Cambridge University
Press, Cambridge, 1992).

\bibitem {Bad} {\it Models of the Nucleon\/} By R.~K.~Bhaduri, (Addison
Wesley, Reading, 1988).

\bibitem {RG} G.~Altarelli, Phys. Rep. {\bf 81} (1982) 1.

\bibitem{Ji95}P.~Hoodbhoy, X.~Ji, and J.~Tang,
MIT-CTP-2476, hep-ph/9510304, 1995.

\bibitem{DGG75} A.~de Rujula, H.~Georgi and
S.~L.~Glashow, Phys. Rev. {\bf D12} 147 (1975).

\bibitem{DJJK75} T.~A.~DeGrand, R.~L.~Jaffe,
K.~Johnson, and J.~Kiskis, Phys. Rev. {\bf D12}
2060 (1975).

\bibitem{GHJ} S.~N.~Goldhaber, T.~H.~Hansson, and R.~L.~Jaffe,
Phys. Lett. {\bf 131B} (1983) 445, Nucl. Phys. {\bf B277} (1986) 674

\bibitem{KS74} J.~Kogut and L.~Susskind, Phys. Rev. {\bf D11}, 3594 (1974).

\bibitem{JK95} J.~Kodaira, Hiroshima University preprint, HUPD-9405 (January
1995), hep-ph 9501381, and references therein.

\bibitem {BL} S.~J.~Brodsky and G.~P.~Lapage, Phys. Rev. {\bf D22}, 2157
(1980).

\bibitem {Bur} M.~Burkardt, Nucl. Phys. {\bf B373} (1992) 371.

\bibitem{RLJ75} R.~Jaffe, Phys. Rev. {\bf D11} (1975) 1953; R.~Jaffe and
X.~Ji, Phys. Rev. {\bf D43} (1991) 724; M.~Stratmann, Z.~Phys. {\bf C60}
(1993), 763.

\bibitem{GRV}M.~Gl\"uck, E.~Reya, and A.~Vogt, Z.~Phys. {\bf C67} (1995) 433,
and references therein.

\end{references}
\end{document}